\documentclass[manuscript]{acmart}
\AtBeginDocument{%
  }

\usepackage{hyperref}
\usepackage{amsmath}
\usepackage[export]{adjustbox}
\usepackage{threeparttable} 
\usepackage{array}
\usepackage{booktabs}
\usepackage{longtable}
\usepackage{makecell}
\usepackage{lscape}
\usepackage{tabularray}

\renewcommand\keywords[1]{}

\setcopyright{none} 
\copyrightyear{}
\acmYear{}
\acmDOI{}

\acmJournal{toit}
\acmVolume{1}
\acmNumber{1}
\acmArticle{1}
\acmMonth{1}

\begin{document}

\title{Shared Channel Capacity and Node Lifetime: An Empirical Study of the Lightning Network}

\author{Danila Valko}
\email{danila.valko@offis.de}
\orcid{0000-0002-8058-7539}
\affiliation{%
  \institution{Carl von Ossietzky Universität Oldenburg}
  \city{Oldenburg}
  \country{Germany}
}
\affiliation{%
  \institution{OFFIS – Institute for Information Technology}
  \city{Oldenburg}
  \country{Germany}
}

\author{Jorge Marx Gómez}
\email{jorge.marx.gomez@uol.de}
\orcid{0000-0002-7833-7549}
\affiliation{%
  \institution{Carl von Ossietzky Universität Oldenburg}
  \city{Oldenburg}
  \country{Germany}
}

\renewcommand{\shortauthors}{Valko et al.}
\acmArticleType{Research}

\begin{abstract}
The Lightning Network (LN) is a rapidly evolving payment channel network that enables scalable, off-chain transactions on top of Bitcoin. While prior research has documented its topological structure and liquidity concentration, the joint relationships between node lifetime, connectivity, and capacity remain insufficiently understood. This study provides a comprehensive empirical analysis of these relationships using a large-scale dataset of LN topology snapshots spanning the period 2019–2023. We examine whether node lifetime influences shared channel capacity, and whether this effect is mediated and moderated by node degree. In addition, we account for hierarchical geographic structure and explore the role of country-level economic conditions. The results show that node lifetime has a positive but relatively modest direct effect on capacity. This relationship is largely mediated by node degree, indicating that liquidity accumulation primarily occurs through increased connectivity. Furthermore, the interaction between lifetime and degree reveals significant heterogeneity, with stronger effects observed among highly connected and high-capacity nodes. Mixed-level models demonstrate superior explanatory power, highlighting the importance of country- and region-level variation. The inclusion of GDP per capita confirms that broader economic conditions significantly influence capacity distribution. Overall, the findings suggest that liquidity in the LN emerges from the interplay of temporal dynamics, network structure, and economic context. This study contributes to a more integrated understanding of payment channel networks and provides a foundation for future research on their evolution and efficiency.
\end{abstract}
\acmCodeLink{}
\acmDataLink{}
\acmContributions{DV and JMG designed the study; DV conducted the experiments and analysed the results, all authors participated in writing the manuscript.}




\maketitle

\section{Introduction}

The Lightning Network (LN) has emerged as a prominent second-layer solution designed to address scalability limitations of the Bitcoin blockchain by enabling fast and low-cost off-chain transactions\cite{R040}. By relying on a network of payment channels, the LN allows participants to route payments through interconnected nodes without recording every transaction on-chain. As a result, the efficiency and reliability of the network depend critically on its structural and economic properties, including node connectivity, liquidity allocation, and participation dynamics\cite{R273, Divakaruni2023}.

A growing body of research has documented that the LN exhibits highly uneven structural characteristics, such as heavy-tailed degree and capacity distributions, as well as substantial churn in node participation\cite{R202, R337}. While these studies provide valuable descriptive insights, they largely focus on individual aspects of the network in isolation. In particular, the joint relationship between node lifetime, connectivity (degree), and liquidity (capacity) remains insufficiently understood. This gap is especially relevant, as these characteristics are likely interdependent: nodes that remain active for longer may accumulate connections and liquidity over time, while highly connected nodes may benefit disproportionately from network effects.

Furthermore, the LN is not only a technical system but also a socio-economic network embedded in a global environment. Nodes are geographically distributed and operate under heterogeneous economic and infrastructural conditions\cite{R323, Zabka2021}. However, the role of such contextual factors, especially at the country and regional levels, has received limited empirical attention\cite{ValkoMarxGomezReview2025}. Similarly, prior research has rarely explored more complex relationships, such as mediation and moderation effects, which could provide deeper insight into the mechanisms underlying liquidity accumulation.

To address these gaps, this study provides a comprehensive empirical analysis of the relationships between node lifetime, degree, and shared channel capacity in the LN. Specifically, we investigate (i) whether longer node lifetime leads to higher capacity, (ii) whether this relationship is mediated by node degree, and (iii) whether degree also moderates the effect of lifetime on capacity. In addition, we account for hierarchical geographic structure and examine the role of country-level economic conditions.

By combining large-scale longitudinal data with regression, mediation, moderation, and mixed-level modelling approaches, this paper contributes to a more integrated understanding of how temporal, structural, and economic factors jointly shape liquidity distribution in the LN. To situate this analysis within existing research, the following section reviews prior work on the structure, dynamics, and geographic distribution of the LN.

\section{Related work}

A growing body of empirical research investigates the structural and economic properties of the LN, with particular attention to its evolution, topology, and liquidity distribution\cite{ValkoMarxGomezReview2025}. Early measurement studies like~\cite{R033, R040, R156, R064, R212, R155} provide foundational insights into the network’s development, documenting its rapid growth, dynamic participation, and evolving connectivity patterns. These works show that the LN is highly dynamic, with frequent node entry and exit, and emphasize the importance of longitudinal analysis for understanding its behaviour.

A central focus of the related literature is the topological structure and degree distribution of the LN. Studies like~\cite{R039, R202, R337, R182} demonstrate that the network exhibits scale-free characteristics, with a highly skewed degree distribution and the emergence of hub nodes. These hubs play a critical role in maintaining connectivity and routing efficiency but also introduce centralization risks. 

Related to topology, several studies examine capacity distribution or liquidity concentration. For example~\cite{R182, R273, R232, R064}, find that channel capacity is highly concentrated, with a small number of nodes controlling a large share of total liquidity. This unequal distribution has important implications for routing performance, fee markets, and systemic resilience\cite{R155, R272, R073}. Other work, such as~\cite{R014, R358, Guasoni2025} focuses on liquidity management and routing efficiency, but typically treats capacity as a given rather than explaining its determinants\cite{ValkoMarxGomezReview2025}.

Another important strand of research addresses node and channel lifetime as well as network churn. Measurement-based analyses, including~\cite{R282, R337}, show that many nodes and channels are short-lived, while a smaller subset remains active over extended periods. These findings highlight the importance of temporal dynamics in the LN. However, existing studies largely provide descriptive statistics on lifetime and churn, without systematically linking these factors to other node-level outcomes such as degree or capacity.

A smaller but growing literature explores the geographical distribution of LN nodes. Studies such as~\cite{R202, Zabka2021, R124, R323, R064} show that nodes are geographically clustered, often in economically developed regions and urban centres. While these works suggest that infrastructure and economic conditions may influence network participation, there is very limited empirical research directly examining this relationship, for instance, between macroeconomic indicators (e.g., GDP per capita) and node-level variables such as capacity or liquidity allocation\cite{ValkoMarxGomezReview2025}. Most economic analyses of the LN instead focus on micro-level incentives, such as routing fees and profitability, leaving broader economic determinants largely unexplored.

From a methodological perspective, the literature is dominated by descriptive network analysis, simulation, and optimization approaches. There is little evidence of studies employing hierarchical or multilevel modelling to explicitly account for the nested structure of nodes within regions and countries. Geographic heterogeneity is occasionally acknowledged, but rarely modelled in a systematic way that captures variation across multiple levels. For further overview we refer to~\cite{ValkoMarxGomezReview2025}.

Moreover, despite increasing interest in causal mechanisms, there is a notable absence of studies applying mediation or moderation analysis in the context of the LN. Existing work typically focuses on pairwise associations, such as the relationship between topology and performance\cite{R182, R323}, without formally decomposing effects into direct and indirect pathways or examining interaction effects between variables.

In summary, prior research has established that the LN exhibits heavy-tailed degree and capacity distributions, strong centralization tendencies, and highly dynamic participation patterns. However, there remains a significant gap in understanding how key node-level characteristics, particularly lifetime, degree, and shared channel capacity are jointly related. In addition, the roles of hierarchical geographic structure, macroeconomic conditions, and causal mechanisms such as mediation and moderation remain largely unexplored. This study addresses these gaps by providing an integrated empirical analysis that links network structure, temporal dynamics, and economic context.

\section{Research design and hypotheses}

While prior considerations suggest that node lifetime and connectivity may influence node- and network-level capacity in the LN, this study adopts an exploratory perspective. Nevertheless, it is useful to formalize these intuitive expectations as testable hypotheses.

First, we posit that longer-lived nodes have more time to accumulate liquidity. Therefore, node lifetime is expected to have a positive direct effect on shared channel capacity ($H_1$).

Second, this accumulation process likely operates through the formation of new payment channels or the replacement of existing ones with higher-capacity channels. Consequently, node degree is expected to mediate the relationship between lifetime and shared capacity ($H_2$). In other words, longer node lifetime leads to higher degree, which in turn increases capacity.

Finally, node degree may also act as a moderator, affecting the strength or direction of the relationship between lifetime and capacity. Given the preferential attachment tendencies in the LN, where new connections are more likely to be established with already well-connected nodes, degree may alter rather than merely explain the effect of lifetime. Thus, we hypothesize that the impact of lifetime on capacity depends on node degree, implying a significant interaction effect between lifetime and degree ($H_3$).

\begin{figure*}[ht]
\centering
\includegraphics[scale=0.3, center]{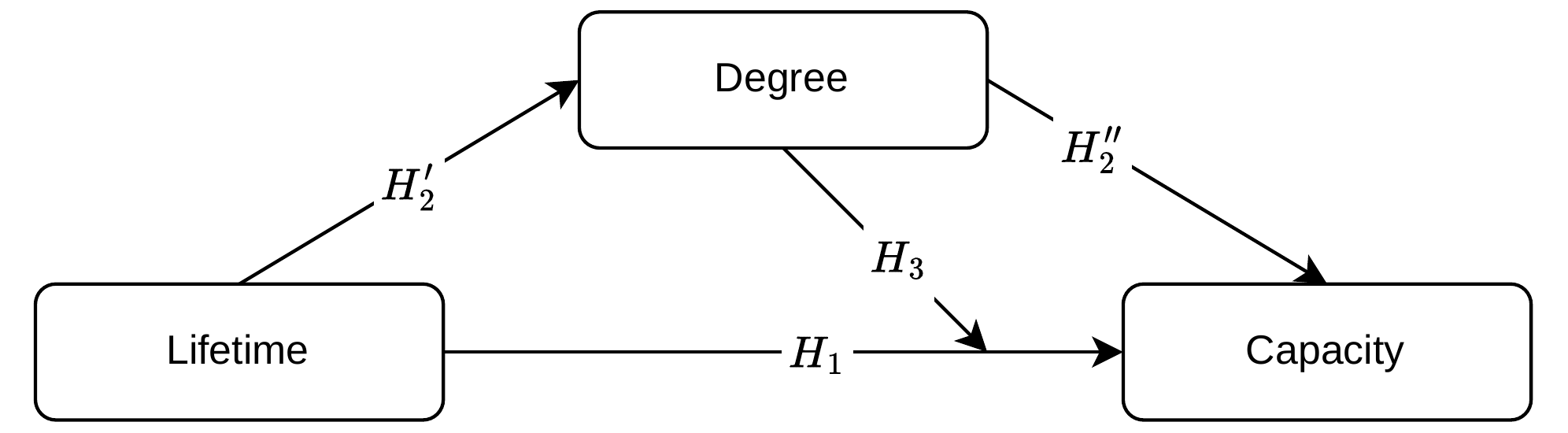}
\caption{Conceptual Model}\label{fig01}
\Description{A conceptual model diagram.}
\end{figure*}

Based on these hypotheses, we define a conceptual model of the tested relationships (Fig.~\ref{fig01}).

\section{Methods}
\subsection{Data source}

This study relies on a public dataset of LN topology snapshots, reconstructed from an archive, comprising more than 35 million LN gossip messages collected between 2018 and 2023 using dedicated LN nodes\cite{lngossip}. The source dataset consists of 336 network snapshots spanning January 2019 to July 2023, providing near-continuous coverage of LN evolution over almost five years\cite{datapaper2025}. To ensure reliability, key network statistics (e.g., number of nodes, channels and network diameter) were benchmarked against independent sources such as \textit{BitcoinVisuals} and \textit{mempool.space}. Statistical tests confirmed that the dataset captures the overall dynamics of the LN with reasonable accuracy despite inherent limitations of observational data, see~\cite{datapaper2025} for further details.

To incorporate spatial information, nodes in the snapshots were geolocated and mapped to country and region based on publicly available IP addresses using an external geocoding service. Geographical coverage is inherently partial due to the widespread use of privacy-preserving technologies such as TOR, which obscure node locations. Over time, the share of geolocatable nodes declines from roughly 40\% to about 15\% with an average of approximately 39\% across the time frame. Nevertheless, validation against external benchmarks showed strong alignment in country-level distributions (with a root mean square deviation for geolocation accuracy of 0.57\% and the Spearman correlation between the snapshots and country-level reference values of 0.98), indicating that the geolocated subsample provides a meaningful approximation of the LN's global spatial structure, for details see \cite{datapaper2025}.

Overall, the source dataset\cite{dataset2025, datapaper2025} represents a large-scale, independently validated source of LN topology and geographic information, enabling the analysis of both network dynamics and cross-country heterogeneity. Based on this dataset, we construct a set of node-level variables capturing connectivity, capacity, and temporal persistence, which are described in the following subsection.

\subsection{Main variables}

This study focuses on three key node-level characteristics of the LN: \textit{Degree}, \textit{Capacity} and \textit{Lifetime}. These variables are derived through a multi-step data preparation pipeline applied to the reconstructed sequence of network topology snapshots referenced above.

\subsubsection{Variable construction pipeline}

We begin by extracting the full list of nodes from each available network snapshot (downloaded from \cite{dataset2025, datapaper2025}), tagging each observation with the corresponding snapshot date. This longitudinal structure allows us to track node activity over time and construct consistent node-level measures.

Node degree is defined as the total number of channels connected to a node in a given snapshot, accounting for both incoming and outgoing connections.

Node capacity is measured as the total liquidity controlled by a node, operationalized as the sum of half the capacity of all channels incident to the node. 
Formally, for a graph snapshot $G=(V,E)$, the shared capacity of node $v \in V$ is defined as:
\begin{equation} \label{eq01}
\begin{split}
Capacity(v) = \frac{1}{2}\sum_{e \in E_v} capacity[e],
\end{split}
\end{equation}
where $E_v \subseteq E$ denotes the set of channels connected to node $v$. This formulation reflects the fact that channel capacity is shared between two endpoints.

Both degree and capacity are first computed at the snapshot level and then averaged for each node across all observed time periods. This averaging step provides stable estimates that mitigate short-term fluctuations in network structure.

Node lifetime is constructed differently, capturing the temporal persistence of a node in the network. Specifically, lifetime is defined as the number of days between the first and the last snapshot in which a node appears. This measure approximates the duration of a node's observable participation in the LN.

\subsubsection{Data filtering and sample construction}

Following variable construction, we apply a series of consistency filters to ensure data quality. We remove nodes with implausible or incomplete characteristics, including nodes with zero degree (1.8\%), zero capacity (6.6\%), or zero lifetime (1.8\%). After filtering, the final dataset consists of 35,481 unique nodes, representing the full analytical sample. Among these, 7,917 nodes are successfully geolocated and assigned to specific countries and regions, while the remaining 27,564 nodes lack geographical information.

It is important to emphasize here that the constructed variables represent \textit{observable proxies} rather than exact measures of underlying node characteristics. Their accuracy is constrained by the temporal coverage and partial observability of the LN, as well as by the snapshot-based reconstruction approach. These limitations are discussed in detail in the corresponding subsection. Nevertheless, the empirical distributions of degree, capacity, and lifetime are consistent with patterns documented in prior studies, providing additional confidence in their validity for analysis.

Finally, all three variables exhibit heavy-tailed distributions (as further demonstrated in the results section) -- a well-known property of financial and network data. To address potential estimation biases, we use logarithmic transformations of these variables in regression analyses. For ease of exposition, however, variable names and model specifications are presented without explicit log notation, unless otherwise stated.

\subsection{Analytical strategy}

\subsubsection{Moderation and mediation analysis}

Moderation and mediation analysis provide complementary perspectives on the relationships between variables. Mediation focuses on identifying the underlying mechanism through which an independent variable affects a dependent variable, by introducing an intermediate (mediating) variable that transmits this effect. In contrast, moderation examines whether the strength or direction of the relationship between an independent and a dependent variable depends on the level of a third variable (the moderator), typically captured through an interaction term.

Formally, mediation is assessed by decomposing the total effect into direct and indirect components, where the indirect effect operates through the mediator. Moderation, on the other hand, is evaluated by including an interaction term between the independent variable and the moderator, allowing the marginal effect to vary across different levels of the moderating variable.

Both mediation and moderation effects can be estimated within a regression framework. In this study, we employ Ordinary Least Squares (OLS) linear models to estimate direct, indirect, and interaction effects. To account for potential heteroscedasticity and the presence of outliers, we additionally estimate Robust Linear Models (RLM) and Quantile Linear Models (QLM), ensuring that the results are not driven by deviations from standard OLS assumptions. Table~\ref{tab01} presents a summary of this analytical strategy, where $\beta_1$, $\beta_2$, and $\beta_3$ represent relationships in question.

\begin{table}[ht]
\centering
\caption{Hypotheses and Corresponding Specifications}
\label{tab01}
\begin{adjustbox}{width=1\textwidth}
\begin{tabular}{p{4.3cm} p{4cm} p{4.8cm}}
\toprule
\textbf{Purpose \& Hypothesis} & \textbf{Relationship} & \textbf{Formula}\\
\midrule

Total effect ($H_1$).
& Lifetime $\rightarrow$ Capacity 
& $\text{Capacity} = \beta_0 + \beta_1 \text{Lifetime} + \epsilon$ \\

Step 1 of mediation analysis ($H_2'$).
& Lifetime $\rightarrow$ Degree 
& $\text{Degree} = \beta_0 + \beta_1 \text{Lifetime} + \epsilon$ \\

Step 2 of mediation analysis ($H_2''$).
& Degree $\rightarrow$ Capacity (controlling for Lifetime) 
& $\text{Capacity} = \beta_0 + \beta_1 \text{Lifetime} + \beta_2 \text{Degree} + \epsilon$ \\


Moderation analysis ($H_3$).
& Lifetime $\times$ Degree $\rightarrow$ Capacity 
& $\text{Capacity} = \beta_0 + \beta_1 \text{Lifetime} + \beta_2 \text{Degree} + \beta_3 (\text{Lifetime} \times \text{Degree}) + \epsilon$ \\


\bottomrule
\end{tabular}
\end{adjustbox}
\end{table}

\subsubsection{Mixed-level analysis}

Since our data includes geolocated nodes, with nodes nested within regions and regions nested within countries, it is reasonable to assume that baseline node capacity and lifetime may vary systematically across these geographical units. To account for this structure, it is possible to (i) simply estimate common regression specifications (see Table~\ref{tab01}) with fixed effects or (ii) use mixed-level models.

The inclusion of fixed effects for countries and regions captures systematic differences in baseline node capacity across geographical units by introducing dummy variables for each country and region (with regions nested within countries). Such fixed-effects account for unobserved heterogeneity across these levels, ensuring that the estimated relationships between node lifetime, degree, and capacity are not confounded by location-specific factors.

By incorporating country and region fixed effects, we can test our hypotheses while controlling for all stable characteristics associated with the nodes' geographic units, see full specifications in the Table~\ref{tab02}.

\begin{table}[ht]
\centering
\caption{Fixed Country and Region Effects Specifications}
\label{tab02}
\begin{adjustbox}{width=1\textwidth}
\begin{tabular}{p{1.8cm} p{5.5cm} p{6.5cm}}
\toprule
\textbf{Hypothesis} & \textbf{Formula} & \textbf{Explanation} \\
\midrule

$H_1$
& $\text{Capacity}_{i} = \beta_0 + \beta_1 \text{Lifetime}_{i} + \gamma_k \text{Country}_k + \delta_{jk} \text{Region}_{jk} + \epsilon_i$ 
& Tests whether lifetime predicts capacity controlling for country ($\gamma_k$) and region ($\delta_{jk}$) fixed effects. $i$, $j$ and $k$ denote nodes ($i$) nested in regions ($j$) nested in countries ($k$).\\

$H_2'$
& $\text{Degree}_{i} = \beta_0 + \beta_1 \text{Lifetime}_{i} + \gamma_k \text{Country}_k + \delta_{jk} \text{Region}_{jk} + \epsilon_i$ 
& Tests whether lifetime predicts degree, controlling for geographic fixed effects. \\

$H_2''$
& $\text{Capacity}_{i} = \beta_0 + \beta_1 \text{Lifetime}_{i} + \beta_2 \text{Degree}_{i} + \gamma_k \text{Country}_k + \delta_{jk} \text{Region}_{jk} + \epsilon_i$ 
& Tests if degree mediates the effect of lifetime on capacity, controlling for geographic fixed effects. \\

$H_3$
& $\text{Capacity}_{i} = \beta_0 + \beta_1 \text{Lifetime}_{i} + \beta_2 \text{Degree}_{i} + \beta_3 (\text{Lifetime}_{i} \times \text{Degree}_{i}) + \gamma_k \text{Country}_k + \delta_{jk} \text{Region}_{jk} + \epsilon_i$ 
& Tests whether the effect of lifetime on capacity depends on degree, controlling for geographic fixed effects. \\

\bottomrule
\end{tabular}
\end{adjustbox}
\end{table}

While this approach captures common fixed effects across observations, it does not explicitly model random variation in intercepts attributable to different hierarchical levels. For this purpose we estimate Mixed-Level (hierarchical) regression Models (MLM), which allow us to include both fixed effects for global predictors and random intercepts for nested groups.

In particular, by including random intercepts for regions nested within countries, we can account for the fact that nodes within the same region or country may share unobserved characteristics differently influencing their capacity (e.g., specific infrastructural, economical or environmental conditions). This approach also enables us to test the hypotheses while controlling for potential clustering effects, see full specifications in the Table~\ref{tab03}.

\begin{table}[ht]
\centering
\caption{Mixed-level Specifications}
\label{tab03}
\begin{adjustbox}{width=1\textwidth}
\begin{tabular}{p{1.8cm} p{5.5cm} p{6.5cm}}
\toprule
\textbf{Hypothesis} & \textbf{Formula} & \textbf{Explanation} \\
\midrule

$H_1$
& $\text{Capacity}_{ijk} = \beta_0 + \beta_1 \text{Lifetime}_{ijk} + u_{0k} + u_{0jk} + \epsilon_{ijk}$ 
& Tests whether lifetime predicts capacity, controlling for random effects at country ($u_{0k}$) and region ($u_{0jk}$) levels. $ijk$ denotes nodes ($i$) nested in regions ($j$) nested in countries ($k$). \\

$H_2'$
& $\text{Degree}_{ijk} = \beta_0 + \beta_1 \text{Lifetime}_{ijk} + u_{0k} + u_{0jk} + \epsilon_{ijk}$ 
& Tests whether lifetime predicts degree while accounting for hierarchical structure. \\

$H_2''$
& $\text{Capacity}_{ijk} = \beta_0 + \beta_1 \text{Lifetime}_{ijk} + \beta_2 \text{Degree}_{ijk} + u_{0k} + u_{0jk} + \epsilon_{ijk}$ 
& Tests if degree mediates the effect of lifetime on capacity with hierarchical structure. \\

$H_3$
& $\text{Capacity}_{ijk} = \beta_0 + \beta_1 \text{Lifetime}_{ijk} + \beta_2 \text{Degree}_{ijk} + \beta_3 (\text{Lifetime}_{ijk} \times \text{Degree}_{ijk}) + u_{0k} + u_{0jk} + \epsilon_{ijk}$ 
& Tests whether the effect of lifetime on capacity depends on degree while accounting for country and region random effects. \\

\bottomrule
\end{tabular}
\end{adjustbox}
\end{table}

\subsubsection{Robustness to country-level conditions}

As noted above, country-specific economic, infrastructural, and institutional conditions may influence both the development of the LN and the characteristics of individual nodes. In particular, macroeconomic factors such as economic growth or downturns may affect the amount of available liquidity, while infrastructural and regulatory environments may shape node lifetime and participation dynamics.

While fixed and random effects model specifications account for unobserved heterogeneity across geographical units, they do not explicitly incorporate measurable country-level characteristics. Including observable proxies for such conditions may therefore improve model interpretability and provide a more direct test of robustness.

To address this, we incorporate GDP per capita, sourced from the World Bank database\cite{WorldBankGDPpC}, as a proxy for the level of economic, infrastructural and environmental development at the country level. Beyond reflecting economic output, GDP per capita is strongly correlated with broader development indicators such as the Global Competitiveness Index\cite{GCI}, the Human Development Index\cite{HDI}, and the Environmental Performance Index\cite{EPI}, as demonstrated in multiple cross-country studies\cite{Wang2022, Susnik2017}.

To ensure comparability across countries and mitigate the influence of short-term policies and fluctuations, GDP per capita is averaged over the period 2019–2023, consistent with the time-frame of our main variables. This averaged variable is included in selected model specifications and reported separately to assess the stability of the estimated relationships in the presence of an explicit country-level control. 

\subsubsection{Parameter estimation and goodness-of-fit considerations}

All model parameters, with the exception of RLM and QLM, are estimated using maximum likelihood approach. For mixed-effects models, we employ full maximum likelihood estimation to ensure comparability of information criteria across model specifications.

Model fit is primarily evaluated using the Akaike Information Criterion (\textit{AIC}) and the Bayesian Information Criterion (\textit{BIC}), both of which balance model fit and complexity by penalizing the number of estimated parameters. Lower values of these criteria indicate a better trade-off between goodness-of-fit and parsimony.

Where applicable, we additionally report coefficients of determination ($R^2$). Depending on the model class, these include standard $R^2$ for OLS models, pseudo-$R^2$ measures for RLM and QLM, and conditional $R^2$ for mixed-effects models. In the latter case, the conditional $R^2$ represents the variance explained by the full model, including both fixed and random effects. These $R^2$-type measures should be interpreted with caution, as they are not directly comparable across different model specifications and are most informative when used to compare models of the same type and structure.

RLMs are estimated using M-estimators\cite{Fan2014}, which do not rely on a likelihood-based framework. Consequently, likelihood-based fit statistics such as \textit{AIC} and \textit{BIC} are not defined for these models. To ensure comparability across all model types, we therefore report scale-based error metrics, specifically the mean absolute error (\textit{MAE}) and the median absolute deviation (\textit{MAD}). These measures provide robust assessments of predictive accuracy and are less sensitive to outliers.

All computations and analyses are conducted in Python using the \texttt{statsmodels} package \cite{Seabold2010}.

\section{Results}

\subsection{Node lifetime and capacity distribution}

\begin{figure*}[ht]
\centering
\includegraphics[scale=0.55, right]{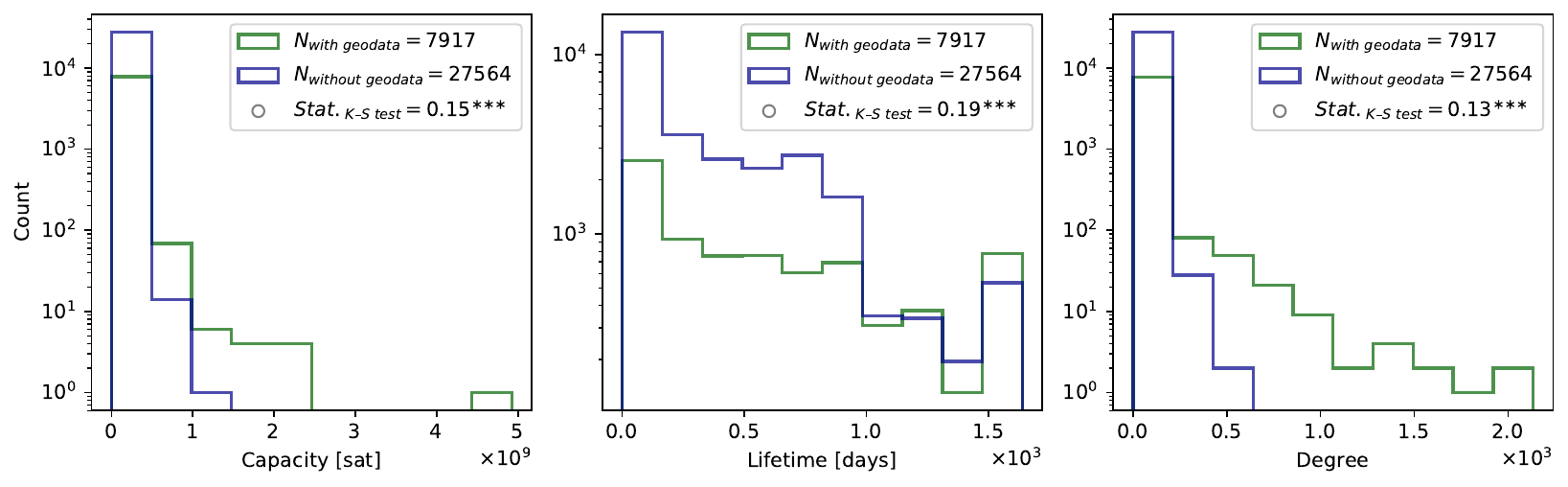}
\Description{Empirical distributions of node capacity, lifetime, and degree for nodes with and without geolocation information.}
\caption{Node Lifetime, Degree and Capacity Distribution}
\label{fig02}

    \begin{tablenotes}
      \small
      \item Note. Significance levels: *** $<$ 0.001, ** $<$ 0.01, * $<$ 0.05, + $<$ 0.1.
    \end{tablenotes}

\end{figure*}

Figure ~\ref{fig02} presents the empirical distributions of node capacity, lifetime, and degree for the two subsamples: nodes with geolocation information and nodes without it. As can be seen, all three variables exhibit strongly right-skewed, heavy-tailed distributions, motivating the use of logarithmic transformations in the subsequent regression analyses.

Left histogram of the Figure~\ref{fig02} shows the distribution of node capacity (in satoshis). The majority of nodes operate with relatively low shared channel capacity, while a small number of nodes concentrate extremely large amounts of liquidity, reaching values on the order of $10^9$ satoshis. This pronounced skewness indicates a high level of inequality in liquidity allocation across the network, a characteristic typical for complex networks with preferential attachment mechanisms. The comparison between nodes with and without geodata reveals significant, but only moderate differences, as indicated by the Kolmogorov–Smirnov (K-S) statistic of 0.15, suggesting that the geolocated subsample is representative in terms of capacity distribution.

Middle histogram depicts the distribution of node lifetime (in days). Similar to capacity, lifetime is unevenly distributed: most nodes exhibit relatively short operational periods, while a smaller subset remains active for substantially longer durations (up to approximately 1,500 days and more). This pattern aligns with the expectation that node participation in the LN is dynamic. The K-S statistic of 0.19 indicates somewhat larger, yet still moderate, differences between the geolocated and non-geolocated samples, though the overall distributional shape remains comparable.

Right histogram presents the node degree distribution, capturing the average number of channels per node. Again, the distribution is highly skewed, with most nodes maintaining only a limited number of connections, while a few highly connected hubs reach degrees in the order of thousands. This observation is consistent with the hypothesized role of degree as both a mediator and moderator in the relationship between lifetime and capacity. The KS statistic of 0.13 suggests significant, but minimal divergence between the two samples.

Taken together, Figure ~\ref{fig02} confirms two important points for the subsequent analysis. First, the heavy-tailed nature of capacity, lifetime, and degree justifies the log-transformation strategy adopted in the regression models. Second, the relatively small differences between nodes with and without geodata support the use of the geolocated subsample in models incorporating geographic controls (fixed and mixed effects), without introducing substantial distributional bias.

\subsection{Correlation analysis}

\begin{figure*}[ht]
\centering
\includegraphics[scale=0.6, center]{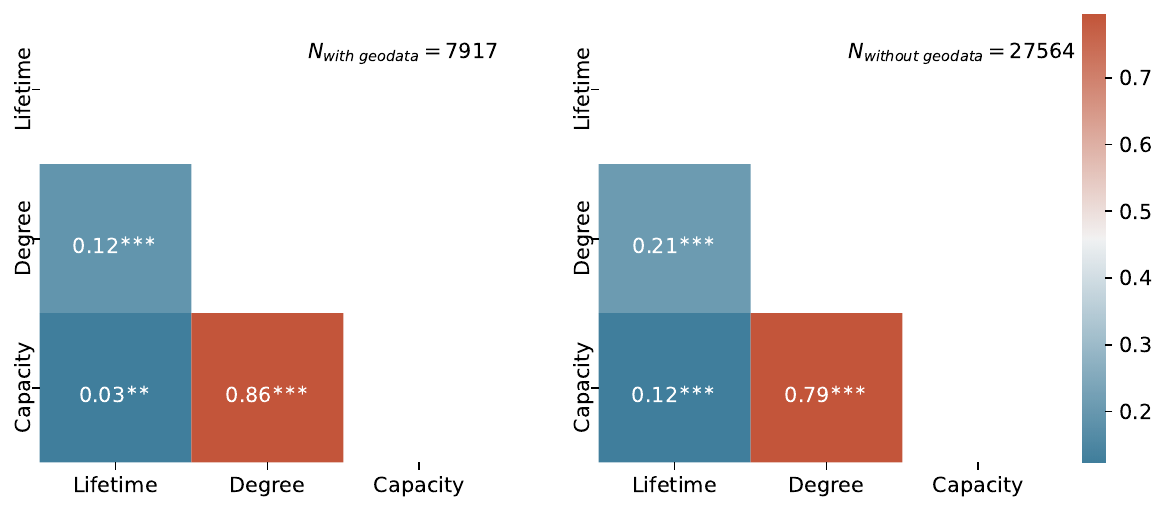}
\Description{Correlation coefficients for node capacity, lifetime, and degree for nodes with and without geolocation information.}
\caption{Node Lifetime, Degree and Capacity Correlation}
\label{fig03}

    \begin{tablenotes}
      \small
      \item Note. Significance levels: *** $<$ 0.001, ** $<$ 0.01, * $<$ 0.05, + $<$ 0.1.
    \end{tablenotes}

\end{figure*}

Figure~\ref{fig03} reports the pairwise correlations between node lifetime, degree, and capacity for both the sample with and without geolocation information. The analysis provides a first descriptive assessment of the relationships formalized above as hypotheses and later tested using regression models. Since the data are highly skewed, we use Spearman's correlation coefficient.

Across both samples, the Spearman correlation coefficient between node degree and capacity is strong and positive (0.86 for nodes with geodata and 0.79 for nodes without geodata). This expectedly indicates that more highly connected nodes tend to control substantially larger amounts of shared channel capacity. The magnitude of this association is markedly higher than for any other pair of variables, providing initial support for the central role of degree in explaining capacity differences and motivating its inclusion as both a mediator and moderator in the analytical framework.

In contrast, the correlation between lifetime and degree is positive but comparatively weak (0.12 with geodata and 0.21 without geodata). This suggests that longer-lived nodes tend to establish more connections, consistent with the accumulation mechanism proposed in $H_2'$, but that lifetime alone explains only a limited share of the variation in node connectivity.

Similarly, the direct correlation between lifetime and capacity is significant, but small (0.03 with geodata and 0.12 without geodata). This weak association indicates that the relationship between lifetime and capacity is unlikely to be purely direct.

Importantly, the correlation structure is consistent across both samples, with only minor differences in magnitude. This further supports the representativeness of the geolocated subsample and justifies its use in subsequent analyses incorporating geographic controls.

Overall, this analysis provides preliminary evidence in line with the proposed hypotheses: while lifetime shows only weak bivariate associations with capacity, its relationship with degree and the strong link between degree and capacity suggest that indirect and interaction effects are likely to play a key role in explaining node-level capacity in the LN.

While these descriptive patterns suggest some associations, we formally test these relationships using regression models presented below.

\subsection{Regression analysis}

As noted above, the variables exhibit heavy-tailed distributions, which may affect regression estimates, we use their log-transformed values in the reported analyses, as noted where appropriate. However, for clarity, the model specifications in the following sections are presented without the logarithmic notation.

\subsubsection{$H_1$}

Table~\ref{tab04} reports the estimation results for $H_1$, which posits a positive relationship between node lifetime and shared channel capacity. Across all model specifications, the estimated $\beta$ coefficient for lifetime is positive and statistically significant, providing consistent support for the hypothesis.

In the baseline OLS models, lifetime exhibits a positive effect on capacity in both subsamples: the geolocated ($\beta \simeq 0.05, p < 0.05$) and without geolocation information ($\beta \simeq 0.21, p < 0.001$). This indicates that longer-lived nodes tend to accumulate greater channel capacity, consistent with the expectation that extended participation allows nodes more time to establish and rebalance channels. However, the effect size is notably smaller in the geolocated subsample, suggesting that part of the variation captured in the ungeolocated sample may be related to unobserved heterogeneity.

To address potential violations of OLS assumptions, RLM are estimated. The results remain substantively unchanged, with lifetime retaining a positive and significant effect of similar magnitude. This confirms that the observed relationship is not driven by outliers or heteroscedasticity, which is particularly relevant given the heavy-tailed distributions discussed above.

Further, models incorporating country and region fixed effects continue to yield positive and significant coefficients for lifetime. This suggests that the relationship between lifetime and capacity is not solely attributable to location-specific factors, but persists even after controlling for systematic differences across geographic units.

Finally, mixed-level models (MLM), which account for hierarchical structure through random intercepts at the country and region levels, also consistently support $H_1$. The lifetime coefficient remains positive and statistically significant, although somewhat attenuated compared to the baseline OLS estimate. At the same time, the reported variance components indicate non-negligible heterogeneity at the country and region levels, justifying the multilevel specification.

Despite the consistent statistical significance, the explanatory power of lifetime alone is limited. The reported $R^2$ values are low across all specifications (ranging approximately from 0.001 to 0.11), indicating that lifetime explains only a small fraction of the overall variation in node capacity. This finding aligns with the weak bivariate correlation observed earlier and suggests that additional mechanisms, such as node degree, are necessary to more fully account for capacity differences.

Overall, the results provide robust empirical support for $H_1$: node lifetime has a positive effect on shared channel capacity ($\beta \simeq 0.05 {\text{---}} 2.1$). However, the relatively small effect size and low explanatory power point to a more complex relationship, motivating the mediation and moderation analyses examined in the subsequent sections.

An important additional insight emerges from the comparison of model fit across specifications. The MLM consistently exhibit superior goodness-of-fit, as indicated by lower $AIC$, $BIC$, $MAE$, $MAD$ values compared to both OLS and RLM specifications, alongside higher conditional $R^2$ values (especially with regional variation included). This improvement in explanatory performance suggests that explicitly accounting for the hierarchical structure of the data captures meaningful variation that would otherwise remain unobserved. In particular, the non-negligible variance components at the country and region levels indicate that both node lifetime and capacity systematically differ across geographical contexts. 

\begin{table}[htbp]
\centering
\caption{$H_1$: Lifetime $\rightarrow$ Capacity}
\label{tab04}
\begin{adjustbox}{width=1\textwidth}
\begin{tabular}{lcccccccccc}
\toprule
& \multicolumn{4}{c}{OLS} & \multicolumn{4}{c}{RLM} & \multicolumn{2}{c}{MLM}  \\
& Without geodata & \multicolumn{3}{c}{With geodata} & Without geodata & \multicolumn{3}{c}{With geodata} & \multicolumn{2}{c}{With geodata} \\
\midrule
Lifetime 
& \makecell{0.210*** \\ {[0.189, 0.230]}} 
& \makecell{0.046* \\ {[0.003, 0.088]}} 
& \makecell{0.046* \\ {[0.003, 0.089]}} 
& \makecell{0.077*** \\ {[0.035, 0.119]}} 
& \makecell{0.212*** \\ {[0.191, 0.232]}}
& \makecell{0.047* \\ {[0.004, 0.090]}}
& \makecell{0.047* \\ {[0.004, 0.090]}}
& \makecell{0.080*** \\ {[0.037, 0.122]}}
& \makecell{0.099*** \\ {[0.058, 0.141]}}
& \makecell{0.136*** \\ {[0.096, 0.176]}}
\\
\makecell[l]{Country fixed effects\\\\} &  &  & \checkmark & \checkmark &  &  & \checkmark & \checkmark &  &  \\
\makecell[l]{Region fixed effects\\} &  &  &  & \checkmark &  &  &  & \checkmark  &  &  \\
Country variation &  &  &  &  &  &  &  &  & \makecell{0.105 \\ (0.024)} & 0.000 \\
Region variation &  &  &  &  &  &  &  &  &  & \makecell{0.144 \\ (0.019)} \\
\midrule
$N_{obs.}$ & 27564 & 7917 & 7917 & 7917 & 27564 & 7917 & 7917 & 7917 & 7917 & 7917 \\
$N_{groups}$ &  &  &  &  &  &  &  &  & 97 & 97 \\
$MAE$ & 0.867 & 1.016 & 1.015 & 0.979 & 0.867 & 1.016 & 1.015 & 0.978 & 0.951 & 0.874 \\
$MAD$ & 0.879 & 1.051 & 1.060 & 0.927 & 0.878 & 1.052 & 1.059 & 0.923 & 0.933 & 0.788 \\
$AIC$ & 80273.1 & 25624.5 & 25624.2 & 25312.3 &  &  &  &  & 25105.1 & 24471.1 \\
$BIC$ & 80289.6 & 25638.5 & 25645.1 & 25340.2 &  &  &  &  & 25133.0 & 24506.0 \\
$R^2$ & 0.01 & 0.001 & 0.001 & 0.04 & 0.01 & 0.001 & 0.001 & 0.04 & 0.07 & 0.11 \\
\bottomrule
\end{tabular}
\end{adjustbox}

    \begin{tablenotes}
      \small
      \item Note. All variables are log-transformed. For clarity, only variables of interest are reported. Significance levels: *** $<$ 0.001, ** $<$ 0.01, * $<$ 0.05, + $<$ 0.1.
    \end{tablenotes}

\end{table}

\subsubsection{$H_2$}

Tables~\ref{tab05} and \ref{tab06} report the results of the mediation analysis, examining whether node degree transmits the effect of lifetime on shared channel capacity. This involves two steps: first, estimating the effect of lifetime on degree ($H_2'$), and second, assessing the effect of degree on capacity while controlling for lifetime ($H_2''$).

In the first step (Table~\ref{tab05}), lifetime has a positive and statistically significant effect on node degree across all model specifications. The estimated coefficients are stable in magnitude ($\beta \simeq 0.09 {\text{---}} 0.13, p < 0.001$), indicating that longer-lived nodes tend to establish more payment channels. This provides strong evidence for the accumulation mechanism: nodes that remain active for longer periods systematically increase their connectivity.

In the second step (Table~\ref{tab06}), degree exhibits a strong positive effect on capacity, while controlling for lifetime. The estimated coefficients for degree are large and highly significant across all models ($\beta \simeq 1.9 {\text{---}} 2.3, p < 0.001$), confirming that more highly connected nodes hold substantially greater shared channel capacity.

Importantly, once degree is included in the model, the coefficient for lifetime becomes negative ($\beta \simeq -0.08 {\text{ to }} -0.15, p < 0.001$). This sign reversal indicates that the positive total effect of lifetime on capacity observed in $H_1$ is largely transmitted through degree, and that the direct effect of lifetime (holding connectivity constant) is negative.

Taken together, these results provide clear evidence of a mediation effect. Lifetime increases node degree, and higher degree in turn leads to greater capacity. The strong magnitude of the degree coefficient, combined with the attenuation and reversal of the lifetime effect in the second step, suggests that degree is a key mechanism linking node longevity to liquidity accumulation in the network.

The robustness of these findings is confirmed across OLS, RLM, and mixed-level specifications. In particular, the mixed-level models again demonstrate improved goodness-of-fit and higher explanatory power (with $R^2$ values reaching approximately 0.7), reflecting the importance of accounting for hierarchical variation across countries and regions in such a complex model. At the same time, the inclusion of geographic effects does not alter the substantive conclusions regarding mediation.

Overall, the results strongly support $H_2$: node degree mediates the relationship between lifetime and capacity ($\beta \simeq 1.9 {\text{---}} 2.3, p < 0.001$). The findings highlight that the accumulation of liquidity is not simply a function of time, but operates primarily through the expansion of network connectivity.

\begin{table}[htbp]
\centering
\caption{$H_2'$: Lifetime $\rightarrow$ Degree}
\label{tab05}
\begin{adjustbox}{width=1\textwidth}
\begin{tabular}{lcccccccccc}
\toprule
& \multicolumn{4}{c}{OLS} & \multicolumn{4}{c}{RLM} & \multicolumn{2}{c}{MLM}  \\
& Without geodata & \multicolumn{3}{c}{With geodata} & Without geodata & \multicolumn{3}{c}{With geodata} & \multicolumn{2}{c}{With geodata} \\
\midrule
Lifetime
& \makecell{0.130*** \\ {[0.123, 0.137]}}
& \makecell{0.099*** \\ {[0.081, 0.118]}}	
& \makecell{0.099*** \\ {[0.081, 0.118]}}	
& \makecell{0.112*** \\ {[0.094, 0.130]}}	
& \makecell{0.127*** \\ {[0.120, 0.134]}}	
& \makecell{0.091*** \\ {[0.073, 0.109]}}	
& \makecell{0.091*** \\ {[0.073, 0.109]}}	
& \makecell{0.101*** \\ {[0.084, 0.119]}}	
& \makecell{0.119*** \\ {[0.100, 0.137]}}	
& \makecell{0.133*** \\ {[0.116, 0.151]}}
\\
\makecell[l]{Country fixed effects\\\\} &  &  & \checkmark & \checkmark &  &  & \checkmark & \checkmark &  &  \\
\makecell[l]{Region fixed effects\\} &  &  &  & \checkmark &  &  &  & \checkmark  &  &  \\
Country variation &  &  &  &  &  &  &  &  & \makecell{0.015 \\ (0.008)} & 0.000 \\
Region variation &  &  &  &  &  &  &  &  &  & \makecell{0.020 \\ (0.006)} \\
\midrule
$N_{obs.}$ & 27564 & 7917 & 7917 & 7917 & 27564 & 7917 & 7917 & 7917 & 7917 & 7917 \\
$N_{groups}$ &  &  &  &  &  &  &  &  & 97 & 97 \\
$MAE$ & 0.298 & 0.425 & 0.425 & 0.413 & 0.296 & 0.423 & 0.423 & 0.409 & 0.403 & 0.375 \\
$MAD$ & 0.373 & 0.542 & 0.541 & 0.512 & 0.353 & 0.522 & 0.523 & 0.482 & 0.492 & 0.420 \\
$AIC$ & 23774.1 & 12402.2 & 12404.0 & 12133.0 &  &  &  &  & 11991.0 & 11475.2 \\
$BIC$ & 23790.5 & 12416.2 & 12424.9 & 12161.0 &  &  &  &  & 12018.9 & 11510.1 \\
$R^2$ & 0.04 & 0.01 & 0.01 & 0.05 & 0.05 & 0.01 & 0.01 & 0.05 & 0.07 & 0.10 \\
\bottomrule
\end{tabular}
\end{adjustbox}

    \begin{tablenotes}
      \small
      \item Note. All variables are log-transformed. For clarity, only variables of interest are reported. Significance levels: *** $<$ 0.001, ** $<$ 0.01, * $<$ 0.05, + $<$ 0.1.
    \end{tablenotes}

\end{table}

\begin{table}[htbp]
\centering
\caption{$H_2''$: Degree $\rightarrow$ Capacity (controlling for Lifetime)}
\label{tab06}
\begin{adjustbox}{width=1\textwidth}
\begin{tabular}{lcccccccccc}
\toprule
& \multicolumn{4}{c}{OLS} & \multicolumn{4}{c}{RLM} & \multicolumn{2}{c}{MLM}  \\
& Without geodata & \multicolumn{3}{c}{With geodata} & Without geodata & \multicolumn{3}{c}{With geodata} & \multicolumn{2}{c}{With geodata} \\
\midrule
Degree
& \makecell{2.255*** \\ {[2.235, 2.274]}}
& \makecell{1.980*** \\ {[1.953, 2.006]}}	
& \makecell{1.979*** \\ {[1.953, 2.005]}}	
& \makecell{1.962*** \\ {[1.936, 1.989]}}	
& \makecell{2.266*** \\ {[2.247, 2.285]}}	
& \makecell{1.996*** \\ {[1.970, 2.022]}}	
& \makecell{1.996*** \\ {[1.970, 2.022]}}	
& \makecell{1.974*** \\ {[1.948, 2.000]}}	
& \makecell{1.952*** \\ {[1.925, 1.978]}}	
& \makecell{1.915*** \\ {[1.887, 1.942]}}
\\
Lifetime
& \makecell{-0.083*** \\ {[-0.096, -0.071]}}
& \makecell{-0.151*** \\ {[-0.173, -0.129]}}
& \makecell{-0.151*** \\ {[-0.173, -0.129]}}
& \makecell{-0.142*** \\ {[-0.165, -0.120]}}
& \makecell{-0.077*** \\ {[-0.090, -0.065]}}
& \makecell{-0.143*** \\ {[-0.165, -0.121]}}
& \makecell{-0.142*** \\ {[-0.164, -0.121]}}
& \makecell{-0.133*** \\ {[-0.155, -0.111]}}
& \makecell{-0.133*** \\ {[-0.155, -0.111]}}
& \makecell{-0.120*** \\ {[-0.142, -0.098]}}
\\
\makecell[l]{Country fixed effects\\\\} &  &  & \checkmark & \checkmark &  &  & \checkmark & \checkmark &  &  \\
\makecell[l]{Region fixed effects\\} &  &  &  & \checkmark &  &  &  & \checkmark  &  &  \\
Country variation &  &  &  &  &  &  &  &  & \makecell{0.017 \\ (0.009)} & \makecell{0.006 \\ (0.008)} \\
Region variation &  &  &  &  &  &  &  &  &  & \makecell{0.019 \\ (0.008)} \\
\midrule
$N_{obs.}$ & 27564 & 7917 & 7917 & 7917 & 27564 & 7917 & 7917 & 7917 & 7917 & 7917 \\
$N_{groups}$ &  &  &  &  &  &  &  &  & 97 & 97 \\
$MAE$ & 0.490 & 0.490 & 0.490 & 0.484 & 0.489 & 0.490 & 0.489 & 0.483 & 0.473 & 0.456 \\
$MAD$ & 0.493 & 0.411 & 0.411 & 0.416 & 0.493 & 0.418 & 0.417 & 0.423 & 0.379 & 0.361 \\
$AIC$ & 50967.9 & 15036.1 & 15033.3 & 14987.4 &  &  &  &  & 14850.1 & 14717.5 \\
$BIC$ & 50992.6 & 15057.1 & 15061.2 & 15022.3 &  &  &  &  & 14884.9 & 14759.4 \\
$R^2$ & 0.66 & 0.74 & 0.74 & 0.74 & 0.67 & 0.76 & 0.76 & 0.76 & 0.74 & 0.74 \\
\bottomrule
\end{tabular}
\end{adjustbox}

    \begin{tablenotes}
      \small
      \item Note. All variables are log-transformed. For clarity, only variables of interest are reported. Significance levels: *** $<$ 0.001, ** $<$ 0.01, * $<$ 0.05, + $<$ 0.1.
    \end{tablenotes}

\end{table}

\subsubsection{$H_3$}

Table~\ref{tab07} presents the results for $H_3$, which tests whether node degree moderates the relationship between lifetime and shared channel capacity. This is evaluated by including an interaction term between lifetime and degree in the regression models.

Across most specifications, the interaction term (Lifetime $\times$ Degree) is statistically significant, indicating that the effect of lifetime on capacity depends on node connectivity. However, both the sign and magnitude of this interaction differ between the subsamples with and without geolocation data.

In models without geolocation information, the interaction term is negative and significant ($\beta \simeq -0.035 \text{ to } -0.046$), suggesting that the marginal effect of lifetime on capacity diminishes as node degree increases. By contrast, in the geolocated subsample, the interaction term is positive and significant in most specifications ($\beta \simeq 0.04 {\text{---}} 0.08$), implying that the effect of lifetime on capacity strengthens for more highly connected nodes. This latter finding is consistent with much of the existing literature and aligns with theoretical expectations derived from the preferential attachment nature of networks.

The negative coefficients observed in the subsample without country and regional geolocation information may reflect systematically different connectivity dynamics over the node lifecycle when such information is not publicly available. Alternatively, they could arise from the unintentional aggregation of heterogeneous nodes located in different geographic contexts. In this case, the result may be driven by a form of Simpson's paradox\cite{Wagner1982}.

Despite this variation, a consistent pattern emerges when considering the other variables. Degree retains a strong positive and highly significant effect on capacity across all specifications ($\beta \simeq 1.7 {\text{---}} 2.4, p < 0.001$), reinforcing its central role as the primary driver of liquidity. At the same time, the coefficient for lifetime remains negative when controlling for degree and the interaction term, consistent with the mediation results above.

The presence of a significant interaction term implies that the relationship between lifetime and capacity cannot be fully captured by a simple additive model. Instead, the effect of node longevity is conditional on the node's position in the network. For low-degree nodes, increases in lifetime may have limited or even negative direct effects on capacity once connectivity is held constant. For high-degree nodes, however, lifetime can amplify capacity accumulation, particularly in the geolocated sample where the interaction effect is positive.

As in previous models, the results are robust across OLS and RLM specifications, and the MLM models again demonstrate superior goodness-of-fit, with lower $AIC$, $BIC$, $MAE$ and $MAD$ values. The inclusion of country and region effects confirms that the moderation effect persists even after accounting for geographic heterogeneity.

Overall, the findings provide support for $H_3$: node degree moderates the relationship between lifetime and capacity. However, the differing signs of the interaction term across samples suggest that this moderation effect might be sensitive to sample composition and potentially influenced by geographic or structural factors. This highlights the importance of considering both mediation and moderation jointly when interpreting the relationship between node lifetime, connectivity, and capacity in the LN.

\begin{table}[htbp]
\centering
\caption{$H_3$: Lifetime $\times$ Degree $\rightarrow$ Capacity}
\label{tab07}
\begin{adjustbox}{width=1\textwidth}
\begin{tabular}{lcccccccccc}
\toprule
& \multicolumn{4}{c}{OLS} & \multicolumn{4}{c}{RLM} & \multicolumn{2}{c}{MLM}  \\
& Without geodata & \multicolumn{3}{c}{With geodata} & Without geodata & \multicolumn{3}{c}{With geodata} & \multicolumn{2}{c}{With geodata} \\
\midrule
Lifetime $\times$ Degree
& \makecell{-0.035* \\ {[-0.069, -0.002]}}
& \makecell{0.075** \\ {[0.032, 0.119]}}
& \makecell{0.076** \\ {[0.032, 0.119]}}
& \makecell{0.073** \\ {[0.030, 0.116]}}
& \makecell{-0.046** \\ {[-0.079, -0.013]}}
& \makecell{0.045* \\ {[0.001, 0.088]}}
& \makecell{0.045* \\ {[0.001, 0.088]}}
& \makecell{0.041$^+$ \\ {[-0.002, 0.084]}}	
& \makecell{0.059** \\ {[0.016, 0.102]}}
& \makecell{0.056* \\ {[0.013, 0.099]}}
\\
Degree
& \makecell{2.336*** \\ {[2.257, 2.416]}}
& \makecell{1.786*** \\ {[1.671, 1.900]}}	
& \makecell{1.785*** \\ {[1.671, 1.900]}}	
& \makecell{1.775*** \\ {[1.661, 1.890]}}	
& \makecell{2.372*** \\ {[2.293, 2.450]}}	
& \makecell{1.881*** \\ {[1.767, 1.995]}}	
& \makecell{1.880*** \\ {[1.766, 1.994]}}	
& \makecell{1.868*** \\ {[1.755, 1.982]}}	
& \makecell{1.800*** \\ {[1.686, 1.913]}}	
& \makecell{1.771*** \\ {[1.658, 1.883]}}
\\
Lifetime
& \makecell{-0.056*** \\ {[-0.085, -0.028]}}
& \makecell{-0.220*** \\ {[-0.266, -0.175]}}
& \makecell{-0.220*** \\ {[-0.266, -0.175]}}
& \makecell{-0.210*** \\ {[-0.255, -0.164]}}
& \makecell{-0.042** \\ {[-0.070, -0.014]}}
& \makecell{-0.184*** \\ {[-0.230, -0.139]}}	
& \makecell{-0.184*** \\ {[-0.229, -0.139]}}
& \makecell{-0.171*** \\ {[-0.217, -0.126]}}
& \makecell{-0.188*** \\ {[-0.233, -0.142]}}
& \makecell{-0.172*** \\ {[-0.217, -0.127]}}
\\
\makecell[l]{Country fixed effects\\\\} &  &  & \checkmark & \checkmark &  &  & \checkmark & \checkmark &  &  \\
\makecell[l]{Region fixed effects\\} &  &  &  & \checkmark &  &  &  & \checkmark  &  &  \\
Country variation &  &  &  &  &  &  &  &  & \makecell{0.017 \\ (0.009)} & \makecell{0.006 \\ (0.008)} \\
Region variation &  &  &  &  &  &  &  &  &  & \makecell{0.019 \\ (0.008)} \\
\midrule
$N_{obs.}$ & 27564 & 7917 & 7917 & 7917 & 27564 & 7917 & 7917 & 7917 & 7917 & 7917 \\
$N_{groups}$ &  &  &  &  &  &  &  &  & 97 & 97 \\
$MAE$ & 0.490 & 0.490 & 0.490 & 0.483 & 0.489 & 0.489 & 0.489 & 0.483 & 0.473 & 0.457 \\
$MAD$ & 0.492 & 0.414 & 0.417 & 0.416 & 0.493 & 0.422 & 0.421 & 0.423 & 0.381 & 0.362 \\
$AIC$ & 50965.6 & 15026.6 & 15023.6 & 14978.6 &  &  &  &  & 14844.7 & 14712.9 \\
$BIC$ & 50998.5 & 15054.5 & 15058.5 & 15020.4 &  &  &  &  & 14886.6 & 14761.7 \\
$R^2$ & 0.66 & 0.74 & 0.74 & 0.74 & 0.67 & 0.75 & 0.76 & 0.76 & 0.74 & 0.74 \\
\bottomrule
\end{tabular}
\end{adjustbox}

    \begin{tablenotes}
      \small
      \item Note. All variables are log-transformed. For clarity, only variables of interest are reported. Significance levels: *** $<$ 0.001, ** $<$ 0.01, * $<$ 0.05, + $<$ 0.1.
    \end{tablenotes}

\end{table}

\subsubsection{Country-level conditions}

Table~\ref{tab08} reports the results of the robustness analysis incorporating GDP per capita as an explicit country-level control. This specification complements the fixed- and mixed-effects approaches by directly accounting for observable macroeconomic differences across countries, rather than relying solely on unobserved heterogeneity captured by country and region effects.

Overall, the inclusion of GDP per capita does not materially alter the main findings. The estimated effect of lifetime on capacity ($H_1$) remains positive and statistically significant across model specifications, with coefficients closely aligned with those obtained in the baseline models. Similarly, the first step of the mediation analysis ($H_2'$) continues to show a stable and positive relationship between lifetime and degree. In the second step of the mediation analysis ($H_2''$), degree retains a strong positive and highly significant effect on capacity, while the coefficient for lifetime remains negative when controlling for degree. This confirms that the mediating role of degree persists even after accounting for differences in economic development across countries. Likewise, the moderation results ($H3$) remain qualitatively unchanged: the interaction term between lifetime and degree continues to be statistically significant in most specifications, supporting the presence of conditional effects.

Importantly, GDP per capita itself exhibits a positive and often statistically significant association with node capacity, particularly in models explaining capacity directly. This suggests that nodes located in more economically developed countries tend to control higher levels of shared channel capacity. Interestingly, GDP per capita has no consistent or significant effect on node degree, indicating that while economic conditions influence the amount of liquidity available, they do not systematically affect the formation of connections.

The stability of the core coefficients across all hypotheses indicates that the previously identified relationships are not driven by omitted country-level economic factors. At the same time, the significance of GDP per capita highlights the relevance of macroeconomic context. Since channel capacity ultimately reflects financial liquidity committed by node operators, it is intuitive that nodes in wealthier economies where access to capital, infrastructure, and financial resources is greater are able to allocate larger amounts of funds to the network.

In sum, incorporating GDP per capita strengthens the interpretation of the results without disrupting the underlying mechanisms. The findings suggest that while node-level dynamics such as lifetime and degree are the primary drivers of capacity, these processes are embedded within broader economic environments that significantly shape the overall distribution of liquidity in the network.

\begin{table}[htbp]
\centering
\caption{Test Country-level Conditions}
\label{tab08}
\begin{adjustbox}{width=1.0\textwidth}
\begin{tabular}{lcccccc}
\toprule
& \multicolumn{2}{c}{OLS} & \multicolumn{2}{c}{RLM} & \multicolumn{2}{c}{MLM}\\
\midrule
& \multicolumn{6}{c}{$H_1$: Lifetime $\rightarrow$ Capacity}\\
\midrule
Lifetime
& \makecell{0.045* \\ {[0.002, 0.088]}}
& \makecell{0.076*** \\ {[0.034, 0.118]}}
& \makecell{0.046* \\ {[0.003, 0.089]}}
& \makecell{0.079*** \\ {[0.037, 0.121]}}	
& \makecell{0.099*** \\ {[0.057, 0.140]}}	
& \makecell{0.138*** \\ {[0.097, 0.178]}}
\\
GDP per capita
& \makecell{0.073 \\ {[-0.042, 0.189]}}
& \makecell{0.189** \\ {[0.076, 0.303]}}	
& \makecell{0.072 \\ {[-0.044, 0.188]}}
& \makecell{0.184** \\ {[0.069, 0.299]}}	
& \makecell{0.231* \\ {[0.002, 0.460]}}
& \makecell{0.239$^+$ \\ {[-0.007, 0.484]}}
\\
Country variation &  &  &  &  & \makecell{0.100 \\ (0.023)} & \makecell{0.089 \\ (0.109)} \\
Region variation &  &  &  &  &  & \makecell{0.115 \\ (0.017)} \\
\midrule
& \multicolumn{6}{c}{$H_2'$: Lifetime $\rightarrow$ Degree}\\
\midrule
Lifetime
& \makecell{0.099*** \\ {[0.081, 0.118]}}
& \makecell{0.112*** \\ {[0.093, 0.130]}}	
& \makecell{0.091*** \\ {[0.073, 0.109]}}	
& \makecell{0.101*** \\ {[0.083, 0.118]}}	
& \makecell{0.118*** \\ {[0.100, 0.136]}}	
& \makecell{0.133*** \\ {[0.115, 0.150]}}
\\
GDP per capita
& \makecell{-0.013 \\ {[-0.063, 0.037]}}
& \makecell{0.034 \\ {[-0.016, 0.083]}}
& \makecell{-0.029 \\ {[-0.077, 0.020]}}
& \makecell{0.015 \\ {[-0.032, 0.062]}}
& \makecell{0.054 \\ {[-0.039, 0.147]}}
& \makecell{0.030}
\\
Country variation &  &  &  &  & \makecell{0.015 \\ {(0.008)}} & 0.000 \\
Region variation &  &  &  &  &  & \makecell{0.020 \\ {(0.003)}} \\
\midrule
& \multicolumn{6}{c}{$H_2''$: Degree $\rightarrow$ Capacity (controlling for Lifetime)}\\
\midrule
Degree
& \makecell{1.980*** \\ {[1.954, 2.006]}}
& \makecell{1.962*** \\ {[1.935, 1.988]}}	
& \makecell{1.997*** \\ {[1.971, 2.023]}}	
& \makecell{1.974*** \\ {[1.948, 2.001]}}	
& \makecell{1.951*** \\ {[1.925, 1.978]}}	
& \makecell{1.914*** \\ {[1.887, 1.942]}}
\\
Lifetime
& \makecell{-0.151*** \\ {[-0.173, -0.129]}}
& \makecell{-0.143*** \\ {[-0.165, -0.121]}}
& \makecell{-0.144*** \\ {[-0.166, -0.122]}}
& \makecell{-0.134*** \\ {[-0.156, -0.112]}}
& \makecell{-0.133*** \\ {[-0.155, -0.111]}}
& \makecell{-0.120*** \\ {[-0.142, -0.098]}}
\\
GDP per capita
& \makecell{0.099** \\ {[0.040, 0.158]}}
& \makecell{0.123*** \\ {[0.064, 0.182]}}	
& \makecell{0.099** \\ {[0.040, 0.157]}}
& \makecell{0.125*** \\ {[0.066, 0.184]}}	
& \makecell{0.119* \\ {[0.015, 0.223]}}
& \makecell{0.111* \\ {[0.018, 0.205]}}
\\
Country variation &  &  &  &  & \makecell{0.016 \\ (0.009)} & \makecell{0.004 \\ (0.008)}\\
Region variation &  &  &  &  &  & \makecell{0.019 \\ (0.008)}\\
\midrule
& \multicolumn{6}{c}{$H_3$: Lifetime $\times$ Degree $\rightarrow$ Capacity}\\
\midrule
Lifetime $\times$ Degree
& \makecell{0.074** \\ {[0.031, 0.118]}}
& \makecell{0.071** \\ {[0.028, 0.114]}}
& \makecell{0.043* \\ {[0.000, 0.087]}}
& \makecell{0.039$^+$ \\ {[-0.004, 0.082]}}
& \makecell{0.058** \\ {[0.015, 0.101]}}
& \makecell{0.055* \\ {[0.012, 0.098]}}
\\
Degree
& \makecell{1.789*** \\ {[1.675, 1.904]}}	
& \makecell{1.779*** \\ {[1.665, 1.894]}}	
& \makecell{1.885*** \\ {[1.771, 1.999]}}	
& \makecell{1.874*** \\ {[1.761, 1.988]}}	
& \makecell{1.803*** \\ {[1.689, 1.916]}}	
& \makecell{1.774*** \\ {[1.661, 1.887]}}
\\
Lifetime
& \makecell{-0.220*** \\ {[-0.265, -0.174]}}
& \makecell{-0.208*** \\ {[-0.254, -0.163]}}
& \makecell{-0.184*** \\ {[-0.229, -0.139]}}
& \makecell{-0.170*** \\ {[-0.215, -0.125]}}
& \makecell{-0.186*** \\ {[-0.232, -0.141]}}
& \makecell{-0.170*** \\ {[-0.216, -0.125]}}
\\
GDP per capita
& \makecell{0.098** \\ {[0.039, 0.157]}}
& \makecell{0.122*** \\ {[0.063, 0.182]}}	
& \makecell{0.098** \\ {[0.040, 0.157]}}
& \makecell{0.125*** \\ {[0.066, 0.184]}}	
& \makecell{0.118* \\ {[0.014, 0.222]}}
& \makecell{0.110* \\ {[0.017, 0.204]}}
\\
Country variation &  &  &  &  & \makecell{0.016 \\ (0.009)} & \makecell{0.004 \\ (0.008)} \\
Region variation &  &  &  &  &  & \makecell{0.019 \\ (0.008)} \\
\midrule
$N_{obs.}$ & 7893 & 7893 & 7893 & 7893 & 7893 & 7893 \\
$N_{groups}$ &  &  &  &  & 90 & 90 \\
\makecell[l]{Region fixed effects\\} &  & \checkmark &  & \checkmark &  & \\
\bottomrule
\end{tabular}
\end{adjustbox}

    \begin{tablenotes}
      \small
      \item Note. All variables are log-transformed. For clarity, only variables of interest are reported. Significance levels: *** $<$ 0.001, ** $<$ 0.01, * $<$ 0.05, + $<$ 0.1.
    \end{tablenotes}

\end{table}

\subsubsection{Moderation effect heterogeneity}

Given the observed differences in the sign and magnitude of the interaction term (Lifetime $\times$ Degree) across subsamples in the $H_3$ results, we further investigate whether this pattern reflects sampling-specific bias or genuine heterogeneity in the moderation effect. To this end, we estimate quantile regression models (QLM), allowing the effect of the interaction term to vary across the distribution of node capacity.

\begin{figure*}[ht]
\centering
\includegraphics[scale=0.55, right]{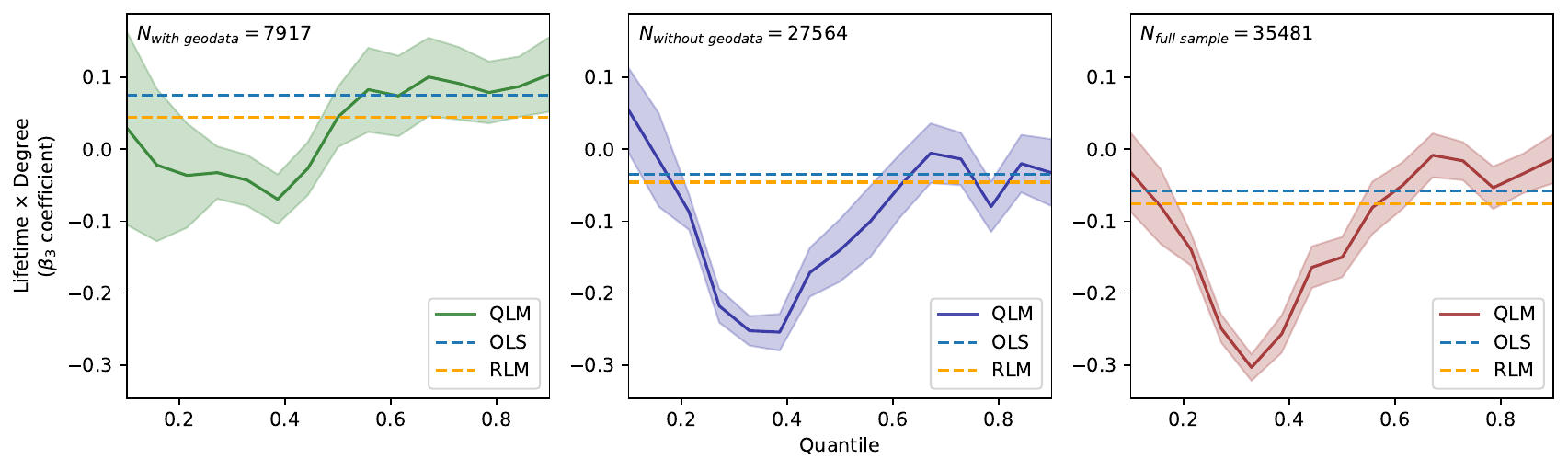}
\Description{Beta coefficients for the interaction term (Lifetime x Degree) using QLM, RLM, and OLS approaches.}
\caption{$\beta_3$ coefficients for Lifetime $\times$ Degree term using QLM, RLM, and OLS.}
\label{fig04}

    \begin{tablenotes}
      \small
      \item Note. All variables are log-transformed. The shaded area beneath the fitted curves represents the 95\% confidence intervals. 
    \end{tablenotes}

\end{figure*}

Specifically, we estimate QLM for deciles between the 10th and 90th percentiles and compare these estimates to the corresponding OLS and RLM point estimates. All standard errors are obtained via bootstrapping ($n_{bootstrap} = 1000$. Figure~\ref{fig04} visualizes the results separately for the geolocated subsample (left panel), the non-geolocated subsample (middle panel), and the full sample (right panel).

The figure reveals substantial heterogeneity in the interaction effect across both outcome distributions and subsamples. In the geolocated subsample, the interaction coefficient varies between approximately $-0.07$ and $0.10$ across quantiles. While lower quantiles exhibit near-zero or slightly negative effects, the coefficient increases steadily toward the upper tail of the capacity distribution, becoming positive and more pronounced. Both OLS and RLM estimates in this subsample are positive, broadly aligning with the upper-quantile behaviour captured by the QLM estimates.

In contrast, the non-geolocated subsample displays a wider and more negative range, with QLM estimates spanning from approximately $-0.25$ to $0.05$. Here, the interaction effect is predominantly negative across most of the distribution, although it approaches zero and in some cases at higher quantiles slightly positive values. Accordingly, both OLS and RLM estimates are negative, reflecting the dominance of lower-quantile effects in this subsample.

For the full sample, the interaction term remains consistently negative across nearly all quantiles, with estimates ranging from approximately $-0.30$ to $-0.01$. However, a clear upward trend is again visible: the magnitude of the negative effect diminishes toward higher quantiles, approaching zero at the upper end of the capacity distribution. The OLS and RLM estimates fall within this range and reflect an average effect that masks this distributional variation.

Taken together, these results indicate that the moderation effect of degree on the lifetime–capacity relationship is not constant but varies systematically across the distribution of capacity. In particular, the interaction effect becomes less negative and in some cases positive as capacity increases. Around the middle of the distribution, the effect begins to shift upward, suggesting that for higher-capacity nodes, longer lifetime combined with greater connectivity may reinforce capacity accumulation, whereas for lower-capacity nodes, the direct effect remains weak or negative when conditioning on degree.

Importantly, this pattern is qualitatively similar across subsamples despite differences in levels, indicating that the apparent inconsistency between OLS and RLM estimates is not merely a sampling artifact. Rather, it reflects underlying heterogeneity in the data that is obscured by OLS/M-based estimators. The quantile regression results therefore provide a more nuanced interpretation of $H_3$: the moderating role of degree is conditional not only on connectivity but also on the position of nodes within the capacity distribution.

\section{Discussion}

\subsection{Contribution and implications}

This study makes several contributions to the growing literature on the LN by providing an integrated empirical perspective on the relationships between node lifetime, degree, and shared channel capacity.

First, the results confirm that node lifetime has a positive effect on capacity, supporting the idea that longer participation enables nodes to accumulate liquidity. This finding aligns with prior studies documenting dynamic participation and churn in the LN, where only a subset of nodes persists over time and contributes more substantially to the network (e.g., \cite{R202, R033}). However, the relatively weak explanatory power of lifetime alone extends previous descriptive findings by showing that temporal persistence is only one part of a broader mechanism.

Second, the mediation analysis highlights node degree as the central mechanism through which lifetime translates into capacity. This result is contributing to the literature on LN topology, which shows that the network exhibits scale-free properties and preferential attachment dynamics (e.g., \cite{R039, R040}). The strong association between degree and capacity also aligns with prior evidence of liquidity concentration among highly connected hub nodes (e.g., \cite{R064, R073}). Our findings extend this line of research by formally demonstrating that connectivity is not only correlated with capacity but also acts as a transmission channel linking temporal dynamics to liquidity accumulation.

Third, the moderation analysis reveals that the relationship between lifetime and capacity is heterogeneous and depends on node degree. This result complements existing studies on LN heterogeneity and centralization (e.g., \cite{R073}), suggesting that highly connected nodes benefit disproportionately from longer participation. The observed variation across quantiles further indicates that these effects are particularly pronounced among high-capacity nodes, reinforcing the notion of cumulative advantage in network growth.

Another key contribution lies in incorporating geographic structure and economic context. The superior performance of mixed-level models demonstrates that country- and region-level factors explain a meaningful share of variation in node characteristics. This finding extends prior research showing that LN nodes are geographically clustered in economically developed regions. Moreover, the significance of GDP per capita supports the interpretation that liquidity allocation is influenced by broader economic conditions, complementing earlier work linking infrastructure and development to network participation\cite{Divakaruni2023}.

Overall, this study advances the literature by moving beyond descriptive analysis toward an explanatory framework that integrates temporal, structural, and economic dimensions. From a practical perspective, these findings have implications for network design and policy. Understanding that liquidity accumulation depends on both connectivity and economic context can inform routing strategies, fee mechanisms, and decentralization efforts aimed at improving the robustness and efficiency of the LN.

From a practical perspective, the results offer actionable insights for both node operators and broader ecosystem governance within the LN. For node operators, the findings suggest that effective node life cycle strategies should prioritize not only longevity but, more importantly, the timely expansion of connectivity. Since capacity accumulation is largely mediated through degree, early-stage nodes can accelerate their growth by strategically forming channels with well-connected peers, rather than relying solely on gradual organic development. This implies a more proactive approach to channel management, including periodic rebalancing and selective link formation to improve routing relevance and fee generation potential. At the same time, the observed heterogeneity indicates that maintaining high connectivity over time is critical for sustaining competitive advantage, particularly for nodes aiming to operate as liquidity hubs.

From a market and policy perspective, the results highlight the importance of economic context in shaping network participation and liquidity distribution. The significance of country-level factors implies that disparities in economic development may translate into structural imbalances within the network. This suggests a role for protocol-level or community-driven mechanisms such as liquidity provisioning incentives, routing fee adjustments, or infrastructure support to encourage broader geographic participation and reduce concentration risks. In this sense, liquidity in the LN can be understood not only as a technical resource but also as an economic one, shaped by access to capital and local conditions. Designing policies that lower entry barriers and support emerging participants could therefore contribute to a more balanced, resilient, and globally distributed network.

\subsection{Limitations}

While these contributions provide new insights into LN dynamics, several limitations should be acknowledged and are discussed below. 

First, the analysis relies on observational data derived from reconstructed snapshots of the LN. Due to the decentralized and partially observable nature of the network, no dataset can capture its full topology at any given point in time. Gossip message propagation delays and independent data collection introduce potential inconsistencies. To address this, the reconstruction pipeline applied chronological ordering, deduplication, and consistency checks, while overlapping snapshots were resolved by retaining the most structurally complete graphs. In addition, key network statistics were systematically validated against independent benchmarks, ensuring that the reconstructed data closely track observed network dynamics.

Second, the main variables -- degree, capacity, and lifetime are constructed as observable proxies rather than exact measures. For example, channel capacity reflects publicly announced values rather than real-time available liquidity, and node lifetime is inferred from first and last observed appearances. To mitigate these limitations, variables were averaged across multiple snapshots, reducing sensitivity to short-term fluctuations and missing observations. Furthermore, nodes with inconsistent or incomplete characteristics (e.g., zero degree, zero capacity, or undefined lifetime) were removed to improve overall data quality and reliability.

Third, geographical coverage is inherently incomplete. An increasing share of LN nodes relies on privacy-preserving technologies such as TOR, which obscure IP addresses and prevent geolocation. This may introduce selection bias in country-level analyses. To manage this issue, the study explicitly reports the share of geolocatable nodes over time and uses the data source validated against external benchmarks. 

Fourth, country-level conditions are proxied using GDP per capita, which captures only one dimension of economic environment. Other relevant factors such as regulatory frameworks or energy costs are not explicitly included. However, the use of mixed-effects models allows unobserved country- and region-level heterogeneity to be partially absorbed through fixed and random effects, reducing omitted variable bias and improving model robustness.

Finally, the empirical framework is based on statistical associations rather than causal identification. While mediation and moderation analyses help uncover plausible mechanisms linking node lifetime, connectivity, and capacity, they do not establish causality. To strengthen inference, the study employs multiple model specifications, including hierarchical models and alternative controls, and tests the stability of results across them. These steps increase confidence in the robustness of the findings, even though causal interpretation remains limited.

Overall, while these limitations reflect inherent challenges in studying decentralized financial networks, the combination of careful data processing, validation, and robustness checks ensures that the results provide a reliable and informative representation of the LN's structural and economic dynamics.

\section{Conclusion}
This paper examines the relationships between node lifetime, degree, and shared channel capacity in the LN using a large-scale, longitudinal dataset of network snapshots. By combining regression, mediation, moderation, and hierarchical modelling approaches, we provide a comprehensive empirical assessment of how these key node-level characteristics interact.

The results show that while node lifetime has a positive effect on capacity, this relationship is largely mediated by node degree, emphasizing the importance of connectivity in liquidity accumulation. At the same time, the moderation analysis reveals that the effect of lifetime varies across different levels of connectivity and capacity, pointing to significant heterogeneity in network dynamics. Furthermore, the inclusion of geographic structure and economic indicators demonstrates that node behaviour is shaped not only by internal network mechanisms but also by external contextual factors.





\section*{Code and data availability}
The data, pipeline and analysis scripts can be accessed via the public repository: \url{https://github.com/ellariel/ln-node-lifetime-analysis/tree/dev}.

\section*{Using generative AI and AI-assisted technologies}
During the preparation of this work, the authors used DeepL to support and refine the grammar and language of the manuscript. The tool was employed solely for linguistic improvements. All intellectual content, ideas, analysis, and conclusions presented in this work remain the sole responsibility of the authors.

\section*{Conflict of interest}
The authors declare that they have no known competing financial interests or personal relationships that could have appeared to influence the work reported in this paper.

\bibliographystyle{ACM-Reference-Format}
\bibliography{_references}

\end{document}